\documentclass[longbibliography,twocolumn,prl,aps,superscriptaddress,showpacs,amsmath,amssymb,floatfix]{revtex4-1}                                                                                                 
\usepackage{graphicx}
\usepackage{amssymb}
\usepackage{amsmath}
\usepackage{epsfig}
\usepackage{color}
\usepackage{tabu}
\usepackage{mathtools}
\usepackage[colorlinks,linkcolor=blue,anchorcolor=blue,citecolor=blue,urlcolor=blue]{hyperref}
\usepackage{physics}
\usepackage{float}
\usepackage{diagbox}

\begin{document}

\title{Anomalous relaxation and multiply time scales in the quantum $XY$ model with boundary dissipation}
\author{Shun-Yao Zhang}
\affiliation{CAS Key Laboratory of Quantum Information, University of Science and Technology of China, Hefei, 230026, People’s Republic of China}
\author{Ming Gong}
\thanks{Email: gongm@ustc.edu.cn}
\affiliation{CAS Key Laboratory of Quantum Information, University of Science and Technology of China, Hefei, 230026, People’s Republic of China}
\affiliation {Synergetic Innovation Center of Quantum Information and Quantum Physics, University of Science and Technology of China, Hefei, Anhui 230026, China}
\affiliation{CAS Center For Excellence in Quantum Information and Quantum Physics}
\author{Guang-Can Guo}
\affiliation{CAS Key Laboratory of Quantum Information, University of Science and Technology of China, Hefei, 230026, People’s Republic of China}
\affiliation {Synergetic Innovation Center of Quantum Information and Quantum Physics, University of Science and Technology of China, Hefei, Anhui 230026, China}
\affiliation{CAS Center For Excellence in Quantum Information and Quantum Physics}
\author{Zheng-Wei Zhou}
\thanks{Email: zwzhou@ustc.edu.cn}
\affiliation{CAS Key Laboratory of Quantum Information, University of Science and Technology of China, Hefei, 230026, People’s Republic of China}
\affiliation {Synergetic Innovation Center of Quantum Information and Quantum Physics, University of Science and Technology of China, Hefei, Anhui 230026, China}
\affiliation{CAS Center For Excellence in Quantum Information and Quantum Physics}
\date{\today}

\date{\today}

\begin{abstract}
	The relaxation of many-body system is still a challenging problem that has not well been understood. In this work we exactly calculate the 
	dynamics of the quantum $XY$ model with boundary dissipation, in which the density matrix in terms of Majorana operators can be decoupled into independent subspaces 
	represented by different number of Majorana fermions. The relaxation is characterized by multiply time scales, and in the long-time limit it is determined by the
	single particle relaxation process in a typical time scale $T^*$. For the bulk bands, we find $T^* \propto  N^3/\gamma n^2 $ in the weak dissipation limit; 
	and $T^* \propto  \gamma N^3/ n^2$ in the strong dissipation limit, where $N$ is the chain length, $\gamma$ is the dissipation rate and $n$ is the band index. 
	For the edge modes $T^* \propto 1/\gamma$, indicating of most vulnerable to dissipation in the long chain limit. 
	These results are counter-intuitive because it means any weak dissipation can induce relaxation, while strong dissipation can induce weak relaxation. We find that these 
	two limits correspond to two different physics, which are explained based on the first and second-order perturbation theory in an equivalent non-Hermitian model. 
	Furthermore, we show that even in the long chain limit the relaxation may exhibit strong odd-even effect. These results shade new insight into the dynamics 
	of topological qubits in environment. 
\end{abstract}
\maketitle

While the dynamics of qubits in environment has been well studied \cite{gardiner1991quantum, johansson2012qutip}, the same issue in the many-body systems 
is still one major challenge in theory \cite{weimer2015variational, le2013steady,karevski2013exact,prosen2014exact,prosen2011exact,rowlands2018noisy,banchi2017driven, medvedyeva2016exact,ribeiro2019integrable,cai2013algebraic,vznidarivc2015relaxation,vznidarivc2011solvable,vznidarivc2010exact} due to more expensive computation cost \cite{complicatetag}.  However, this is an important question at least from two diverse aspects. 
The many-body systems may possess some features that are totally different from the single particle systems, such as ergodicity and thermalization 
\cite{eisert2015quantum,nandkishore2015many}, 
which are fundamental concepts in statistics. In the trapped ions, it may exhibit different dynamics depending strongly on the initial states, which are
explained based on quantum many-body scar \cite{bernien2017probing,turner2018weak,turner2018quantum}. Moreover, it is also an important issue in topological quantum computation \cite{nayak2008non,freedman2003topological,sau2010generic}, in which the two ground states are separated from the excited bands by a finite energy gap \cite{zhang2018quantized, nadj2014observation, wang2018evidence, sun2016Majorana, xu2015experimental, liu2018robust, mourik2012signatures, deng2012anomalous, das2012zero}. Thus if the temperature is much lower than the excitation gap, the 
occupation of the excited states are exponentially small. This picture is not necessarily true in the presence of dissipation, which can induce direct coupling between ground states and 
excited states. 

\begin{figure}
    \centering
    \includegraphics[width=0.48\textwidth]{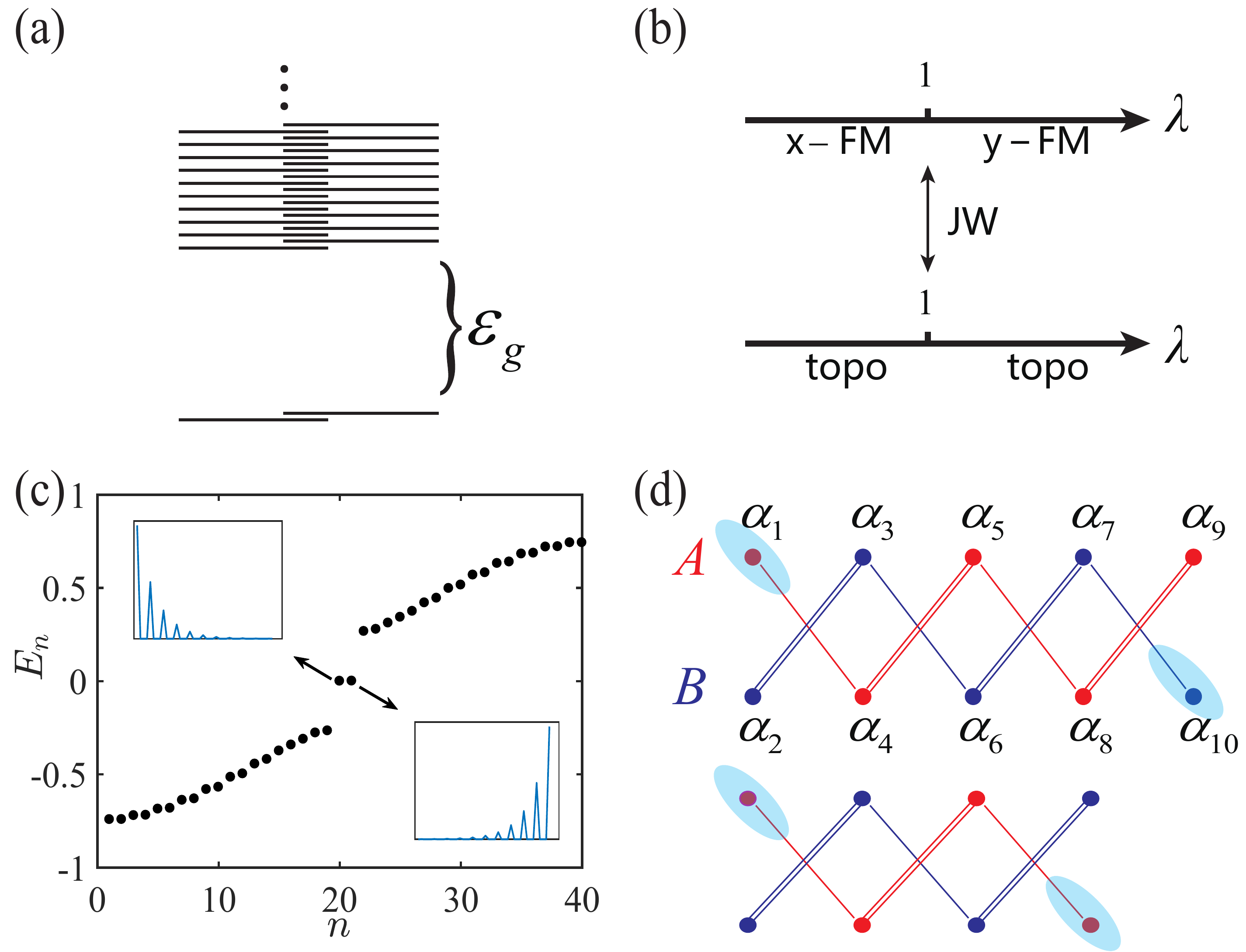}
	\caption{(a) The two lowest energy levels in the $XY$ model are separated from the excited bands by an energy gap $\epsilon_g$. 
	(b) The two ferromagnetic (FM) phases in the $XY$ model will be mapped to two distinct topological $p$-wave superconducting phases after fermionization. 
	(c) Energy levels in the fermion representation, in which the two edge modes, occupied or unoccupied, give the two-fold degeneracy of the $XY$ model. 
	(d) The odd-even effect due to the oscillation of coupling between the two edge modes in the two subchains $A$ and $B$. }
    \label{fig-fig1}
\end{figure}

Here  we explore the roles of edge modes and bulk bands in the dynamics of the quantum $XY$ model with boundary dissipation, in which the two ground states are protected 
by a finite gap. (I) In terms of Majorana operators, the density matrix is decoupled into different subspaces represented by different number of Majorana fermions. In time evolution
the density matrix exhibits multiply relaxation scales, in which the slowest decay is given by the single particle relaxation. This time scale $T^*$ is used to determine the 
relaxation time of the many-body ground state with dissipation. (II) In the weak dissipation limit, $T^* \propto N^3/\gamma $,
where $N$ is the total chain length and $\gamma$ is the boundary dissipation rate. However, in the strong dissipation limit, $T^* \propto \gamma N^3$. The edge modes
are shown to be most vulnerable to dissipation due to $T^* \propto 1/\gamma$. 
These results are counter-intuitive because it means that weak dissipation can induce fast relaxation, while strong dissipation can induce weak relaxation.
We understand these results by mapping the single particle dynamics to a non-Hermitian model. (III) This dynamics exhibits strong odd-even effect in 
the weak dissipation limit, which reduces to a unified form in the strong dissipation limit. These results shade new insight into the dissipation and relaxation of the 
topological qubits in environment.

{\it Model and Master Equation}. We consider a quantum $XY$ model with boundary dissipation, which reads as 
\begin{equation}
	\dot{\rho} = \mathcal{L}(\rho)=-i[H_\text{XY},\rho] + \mathcal{D}(\rho).
    \label{eq-$XY$all}
\end{equation}
Here $H_\text{XY}= -\sum_{i=1}^{N-1} (g_1\sigma_i^x\sigma_{i+1}^x + g_2 \sigma_i^y\sigma_{i+1}^y)$ \cite{osterloh2002scaling, zhu2006scaling, eisert2006general,osborne2002entanglement, cheng2017scaling, osborne2002entanglement} and the dissipation by the Lindblad operator is \cite{hein2005entanglement, cai2006decoherence} 
$\mathcal{D}(\rho)=\frac{\gamma}{2}\sum_{j=1,N}(2\sigma_j^z \rho \sigma_j^z- 2 \rho)$. 
This model is fermionized using Majorana operators via Jordan-Wigner transformation as
$\alpha_{2j-1}=(\prod_k^{j-1}\sigma_k^z)\sigma_j^x,\alpha_{2j}=(\prod_k^{j-1}\sigma_k^z)\sigma_j^y$, after which
\cite{jordan1928p,sachdev2011quantum,kitaev2001unpaired} 
\begin{equation}
    H=ig_1\sum_{j=1}^{N-1}\alpha_{2j}\alpha_{2j+1} - ig_2\sum_{j=1}^{N-1}\alpha_{2j-1}\alpha_{2j+2}.
	\label{eq-JW}
\end{equation}
In this case, the Lindblad operator is still made by local dissipation \cite{dissipationtag}
\begin{equation}
\mathcal{D}(\rho) = -\gamma\sum_{j=1,N} (\alpha_{2j-1} \alpha_{2j} \rho \alpha_{2j-1} \alpha_{2j} + \rho).
\end{equation}

\begin{figure}
    \centering
    \includegraphics[width=0.48\textwidth]{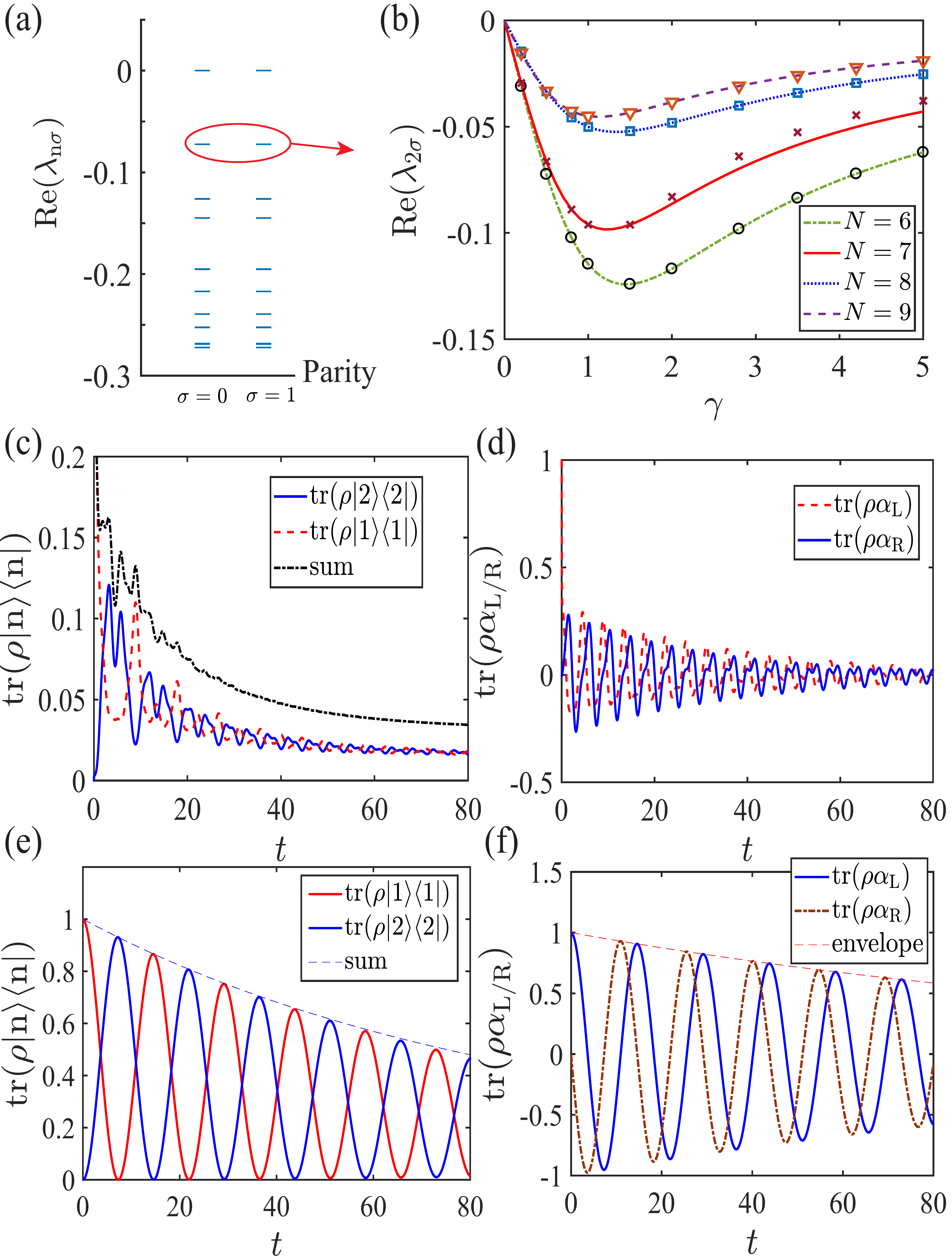}
	\caption{(a) Eigenvalues Re ($\lambda_{n\sigma}$) of the super-operator $\mathcal{L}$ in $XY$ model with $N = 6$, $\gamma = 0.5$.
	Here Re$(\lambda_{n\sigma})$ is arranged in descending order with Re$(\lambda_{n\sigma})$ $\ge$ Re$(\lambda_{n+1\sigma})$, and $\sigma = 0, 1$
	accounts for different parity. 
	(b) The lines are eigenvalues of $H_0 + i\Gamma$ (see Eq. \ref{eq-nH}) with smallest imaginary energy and symbols are Re($\lambda_{2\sigma}$)
	of $\mathcal{L}$. In both figures $g_1 = 1.0$, $g_2 = 0.7$. (c) and (d) show the projection of $\rho(t)$ to the 
	ground state of the $XY$ model $|n\rangle$ ($n =1, 2$); and to the edge modes $\alpha_\text{R/L}$ of the fermion model, respectively, where
	$\rho(0) = |1\rangle \langle 1|$. In both figures $g_1 = 1.0$, $g_2 = 0.7$, $N = 6$ and 
	$\gamma = 20$. (e) and (f) plot the same results as (c) and (d) with $\gamma = 0.005$.}
	\label{fig-fig2}. 
\end{figure}

The spectra between the $XY$ model and $p$-wave superconducting model are related by 
\begin{equation}
E = \sum _i n_i \epsilon_i, \quad n_i = \{0, 1\},
	\label{eq-Eg}
\end{equation}
thus the $N$ eigenvalues from the single particle Hamiltonian can be used to construct all the $2^N$ eigenvalues in the quantum XY model (see Fig. \ref{fig-fig1} (a) 
and (c)). In the fermion representation the two localized zero modes at the open ends give rise to the two-fold degeneracy in the $XY$ model. We focus
on $g_1 > 0$ and $g_2 > 0$, and the phase diagram for the $XY$ model and its corresponding single fermion phases are given in  Fig. \ref{fig-fig1} (b), with 
boundary at $\lambda = g_2/g_1 = 1$. In fermion representation it can be regarded as two separate Majorana chains $A$($\alpha_1,\alpha_4,\alpha_5,\cdots$) 
and $B$($\alpha_2,\alpha_3,\alpha_6,\cdots$), as shown in Fig. \ref{fig-fig1} (d). We can bring Eq. \ref{eq-JW} to a form of 
paired Majorana operators \cite{kitaev2001unpaired,kitaev2009topological}
\begin{eqnarray}
	H = \frac{i}{2}\sum_{k=1}^{N-1} \epsilon_{k} b_{k}^{\prime} b_{k}^{\prime\prime}+{i\delta E_c \over 2} \alpha_\text{L} \alpha_\text{R}, 
	\label{eq-pairings}
\end{eqnarray}
where $\delta E_c \sim e^{-N/\xi}[(-1)^N+1]$. After a special orthogonal transformation the Hamiltonian in Eq. \ref{eq-JW} can be brought 
into small Jordan blocks, with $b_k'$ and $b_k''$ are some new Majorana operators, following the prominent work by Kitaev \cite{kitaev2001unpaired}.
By a transformation from Majorana fermion to canonical fermion $c_k = (b_k' + ib_k'')/2$ and $c_k^{\dagger} = (b_k' - ib_k'')/2$, the spectra of the bulk bands can be obtained as $\pm \epsilon_{k}$.
In the last term  $\alpha_\text{L}$ and $\alpha_\text{R}$ are edge modes at the left and right edges, which can be written as $\alpha_\text{L} = \alpha_1 - 
\lambda \alpha_5 + \lambda^2 \alpha_9 - \cdots $ and $\alpha_\text{R} = \alpha_{2N} - \lambda \alpha_{2N-4} + \lambda^2 \alpha_{2N-8} - \cdots$ (for $\lambda<1$). 
In $\delta E_c$, $\xi \propto 1/ |\ln g_1 - \ln g_2| $ defines the correlation length \cite{kitaev2001unpaired}. This odd-even effect is a typical feature of 
coupling between two distant zero modes, which may happen even in continuous space \cite{cheng2009splitting}. With these operators, the two ground states in Fig. \ref{fig-fig1} (a)
can be written as 
\begin{equation}
	|1\rangle = {1 + \alpha_\text{L} \over \sqrt{2}} |0\rangle, \quad |2\rangle = {1 - \alpha_\text{L} \over \sqrt{2}} |0\rangle,
\end{equation}
where $|0\rangle$ is the ground state satisfying $c_k |0\rangle = 0$ for all $k$ (thus $|0\rangle = \prod_k c_k|\text{vac}\rangle$, where $|\text{vac}\rangle$ is the vaccum state).

This dynamics respects the parity symmetry $[\mathcal{L},P]=0$ with $P=\prod_j^N \sigma_j^z = i^N\prod_j^{2N}\alpha_{j}$. In the long time
limit,
\begin{equation}
	\bar{\rho} = \lim_{t\rightarrow \infty} \rho(t) = (I + c P)/2^N,
	\label{eq-c}
\end{equation}
where $c = \langle \Psi_0 | P | \Psi_0 \rangle$. This corresponds to the maximal mixed state at infinite high temperature. 

{\it Evolution of the density matrix}. The density matrix can be written as Kronecker product of Pauli matrices. However, it is more convenient  
to write this matrix in terms of Majorana operators as following \cite{calabrese2005evolution, vidal2003entanglement, eisert2010colloquium}, 
\begin{equation}
    \rho = {1\over 2^N}\sum c_{a_1,a_2,\cdots,a_{2N}} \alpha_1^{a_1}\alpha_2^{a_2}\cdots \alpha_{2N}^{a_{2N}},  
\end{equation}
where $a_j=\{0, 1\}$. Thus the dynamics of $\rho$ is decoupled into different subspaces denoted as $\mathcal{K}_n$ 
for $n = 0 - 2N$, where $n$ is the number of Majorana operators given by $n = \sum_i a_i$. Formally, we have
\begin{equation}
	\mathcal{K} = \mathcal{K}_0 \oplus \mathcal{K}_1 \oplus \mathcal{K}_2 \oplus \cdots  \oplus \mathcal{K}_{2N},
\end{equation}
where the dimension of $\mathcal{K}_i$ is $C_{2N}^i$. We can readily check that the dimension of $\mathcal{K}$ (the whole space of the Hamiltonian) is 
$C_{2N}^0 + C_{2N}^1 + C_{2N}^2 + \cdots + C_{2N}^{2N} = 4^N$.
This decoupling is essentially the same as the probability distribution in the classical Ising model by Glauber \cite{glauber1963time}. 
By a direct comparison with Eq. \ref{eq-c}, we can find that while the first term ($a_j \equiv 0$) and the last term ($a_j\equiv 1$) are 
unchanged,  which give Eq. \ref{eq-c}; while all the other terms will disappear in the long time limit. 

In principle, the dynamics of $\rho$ can be calculated using hierarchy equations \cite{glauber1963time}. We may
define the following variables $\psi_i = \text{Tr} (\rho\alpha_i)$, $\psi_{ij} = \text{Tr} (\rho\alpha_i\alpha_j)$, $\psi_{ijk} = \text{Tr} 
(\rho\alpha_i \alpha_j \alpha_k)$ (with $i\ne j\ne k$) and then calculate their time evolution based on the Heisenberg equations. 
We find that for the boundary dissipation in Eq. \ref{eq-$XY$all}, the dynamics of these variables are restricted to their own subspaces 
$\mathcal{K}_{1,2,3}$. This limit will show up the multiply time scales during dynamics much more clearly. Obviously, this approach
can be generalized to models with more complicated dissipation and many-body interaction. 

We first compare the full calculation of $\rho$
against the dynamics in subspace $\mathcal{K}_1$. In Fig. \ref{fig-fig2} (a), we present the real part of the eigenvalues of superoperator $\mathcal{L}$ 
of Eq. \ref{eq-$XY$all}, in which the two zero eigenvalues correspond to the unchanged state in $\mathcal{K}_0$ and $\mathcal{K}_{2N}$ subspaces. 
We also calculate the corresponding eigenvalues of this superoperator in the subspace constructed by $\mathcal{K}_1$, which is given in Fig. \ref{fig-fig2} (b). 
We find that the smallest eigenvalue Re($\lambda_{1\sigma}$) of the $\mathcal{L}$ in the $XY$ model is the same as the spectra in $\mathcal{K}_1$ subspace, indicating 
that in the long time limit, all the higher-order terms in $\mathcal{K}_{k\ge 2}$ subspaces decay much faster than that in $\mathcal{K}_1$, leaving $\mathcal{K}_1$ to be
the dominated relaxation channel for the quantum $XY$ model. In Fig. \ref{fig-fig2} (c) and (d), we show the projection of $\rho(t)$ to the ground states of the $XY$
model and the edge modes. In (c), it will approaches $1/2^N$ ($c = 0$); while in (d), it will approaches zero, as expected.

To reinforce this conclusion, we also calculate the eigenvalues of the superoperator $\mathcal{L}$ in subspaces $\mathcal{K}_2$ and $\mathcal{K}_3$. Let us
denote the eigenvalues as $\lambda$ in each subspace. 
We find that, roughly, the slowest decay rate in $\mathcal{K}_2$ is two times faster than that in $\mathcal{K}_1$. Similarly, the decay rate in 
$\mathcal{K}_3$ is much faster than that in $\mathcal{K}_2$. The similar relations can be found in Ref. \cite{glauber1963time} by a finite truncation of
the hierarchy equations. Thus in the long time and long chain limits, we can fully characterize the relaxation time of the many-body system in terms of single particle
decay rate. 

\begin{figure}
    \centering
    \includegraphics[width=0.48\textwidth]{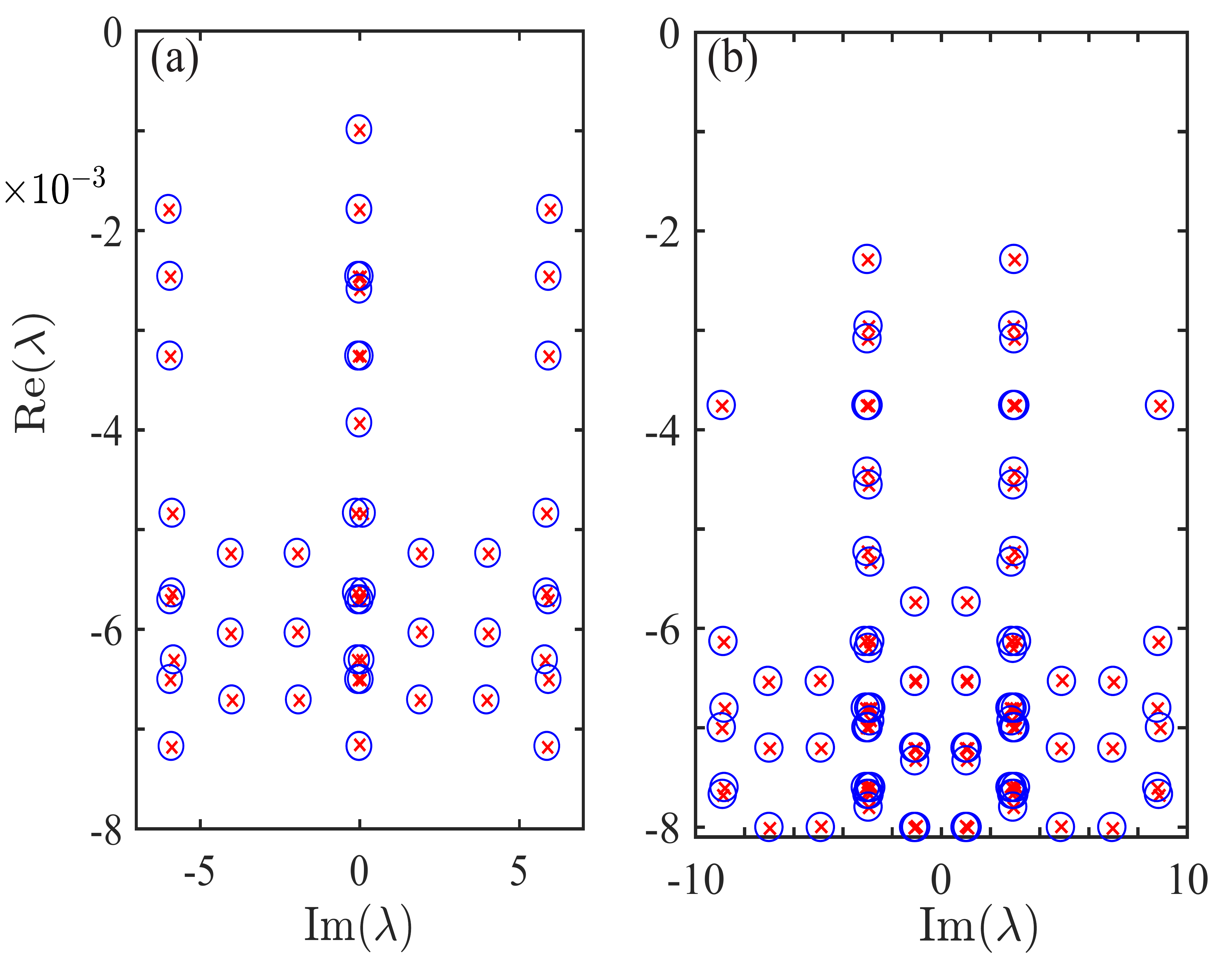}
	\caption{Eigenvalues of the superoperator $\mathcal{L}$ and multiply time scales in the subspaces $\mathcal{K}_2$ (a) and 
	$\mathcal{K}_3$ (b) with $g_1=1,g_2=0.5,\gamma=0.5$ and $N=30$ (by open circles).
	For comparison, we also present $\lambda_i^1 + \lambda_j^1$ in (a) and $\lambda_i^1+\lambda_j^1+\lambda_k^1$ in (b) with crosses, 
	where $\lambda_i^1$ are eigenvalues of $\mathcal{L}$ in the subspace $\mathcal{K}_1$.}
    \label{fig-fig3}
\end{figure}

With this density matrix, we can understand the dynamics in the many-body state for any given initial wave function. For example, for the 
result in Fig. \ref{fig-fig2} (c) and (e), we can express the dynamics of $|n\rangle \langle n|$ as
\begin{eqnarray}
	\text{Tr}(\rho(t) |n\rangle \langle n|) &&  = {1\over 2^N} (c_{0\cdots} + \sum_{i} c_{0\cdots 1_i\cdots 0}(t) \langle n| \alpha_i|n\rangle + \nonumber \\
											&& \sum_{ij} c_{0\cdots 1_i\cdots 1_j \cdots 0}(t) \langle n| \alpha_i \alpha_j |n\rangle + \cdots) +  \\
											&& \sum_{ijk} c_{0\cdots 1_i\cdots 1_j \cdots 1_k \cdots }(t) \langle n| \alpha_i \alpha_j \alpha_k |n\rangle + \cdots), \nonumber
\end{eqnarray}
where the coefficients can be determined by calculation the dynamics of the density matrix in each subspace $\mathcal{K}_i$. Initially, we may find that 
the coefficients $c_{a_1,a_2,\cdots, a_{2N}}$ and the overlap $\langle n|\alpha_1^{a_1} \alpha_2^{a_2} \cdots \alpha_{2N}^{a_{2N}} | n\rangle$ are in the order 
of unity, which after a long time relaxation will finally approach the steady solution in Eq. \ref{eq-c}. For this reason not only the single particle terms, but 
also all the higher order terms, which correspond to the many-body relaxation, will contribute to the relaxation process. 
As a result the dynamics of the many-body state will exhibit multiply time scales during relaxation, in which some of these lifetime scales are shown in 
Fig. \ref{fig-fig3}. Due to the presence of these multi-particle relaxations, we find that in general the decay rate of $|n\rangle \langle n|$ is faster than the 
single particle ones (see Fig. \ref{fig-fig2}). Noticed that in Fig. \ref{fig-fig2} (e) and (f), the decay processes have the same oscillation 
period due to the finite coupling between the two edge modes, which in the many-body case will have the same energy splitting in the two ground state energy 
in the XY model (see Fig. \ref{fig-fig1} (a) and Eq. \ref{eq-Eg}). However, in an odd chain, this oscillation will disappear due to the absence of coupling. 

{\it Relaxation in the long chain limit}. The above results have established a connection between many-body dynamics and single particle dynamics in the long time limit. 
Some more issues need to be explained. (1) Why Re($\lambda_{2\sigma}$) exhibits an inflexion point at $\gamma_c \sim 1$; (2) What will happen in the long chain limit?
and (3) What are the different roles played by the edge modes and bulk modes during relaxation? We focus on subspace $\mathcal{K}_1$, in which the dynamics of $\psi_i$ is given by the 
following non-Hermitian schr\"odinger equation (see Fig. \ref{fig-fig5} (a))
\begin{equation}
    i\partial_t\Psi = 2(H_0 + i \Gamma) \Psi,
	\label{eq-nH}
\end{equation}
where  $\Gamma = \text{diag}(-\gamma,0,\cdots,0,-\gamma)$ and 
\begin{equation}
    H_{0}=\left(\begin{array}{ccccc}
        0 & ig_{2} & 0 & \cdots & 0\\
        -ig_{2} & 0 & ig_{1} & \cdots & 0\\
        0 & -ig_{1} & 0 & \ddots & 0\\
        \vdots & \vdots & \ddots & \ddots & ig_{2}\\
        0 & 0 & 0 & -ig_{2} & 0.
    \end{array}\right).
	\label{eq-H0}
\end{equation}
Here we have defined $\Psi = (\psi_1,\psi_4,\psi_5,\cdots)$ in chain A, while its treatment for chain B is similar. The eigenvalues in this bordered matrix 
are determined by \cite{kouachi2006eigenvalues,yueh2005eigenvalues,willms2008analytic}                                                                         
\begin{eqnarray}
    \Delta_N = \frac{( g_1 g_2 )^{m-1}}{\sin\theta}[ g_1 g_2 (2 i \gamma + \epsilon) \sin(m+1) \theta \nonumber\\
    +(-\gamma^2 \epsilon + i g_1^2\gamma + i g_2^2\gamma)\sin(m\theta)]= 0,
    \label{eq-odd}
\end{eqnarray}         
when $N=2m+1$ is odd; and 
\begin{eqnarray}
    \Delta_N = \frac{( g_1 g_2 )^{m-1}}{\sin \theta}[(-\gamma^2 + g_1^2 + i 2\gamma \epsilon) \sin (m\theta) \nonumber \\
    +  g_1 g_2  \sin(m+1)\theta - \gamma^2 \frac{ g_2}{ g_1 }\sin(m-1)\theta]=0,
	\label{eq-even}
\end{eqnarray}
when $N=2m$ is even, via $\text{Det}(H - \epsilon)$. In the above equations,  
$\epsilon$ is the eigenvalue and its relation to $\theta$ is determined by
\begin{equation}
    \epsilon^2 = g_1^2 + g_2^2 + 2 g_1 g_2 \cos \theta.
	\label{eq-epsilon}
\end{equation}
When $\gamma = 0$ for odd chain, we have $\epsilon_n = \pm  \sqrt{g_1^2 +g_2^2 +2 g_1 g_2  \cos \theta_n }$ and $\theta_n = n\pi/(m+1)$ for $n= 1, \cdots,m$ and $\epsilon_0=0$. 
Thus the energy gap in Fig. \ref{fig-fig1} (a) is given by $\varepsilon_g = \vert g_1 -g_2 \vert$.

We find that for the extended bands, the eigenvalues and the phase can be written as
\begin{equation}
    \epsilon_n = \epsilon_{n,\text{r}} - i \epsilon_{n,	\text{i}}, \ \theta_n = \frac{n \pi}{m+1} + z_{n,\text{r}} + i z_{n,\text{i}},    
    \label{eq-approx}
\end{equation}
where $n \ll m$ and $\epsilon_{n,	\text{i}}$, $z_{n, \text{r}}$, $z_{n,\text{i}}$ are small numbers in the sense that $\lim_{m\rightarrow \infty} m z_{n,i/r} = 0$. 
These solutions can be obtained by linearizing the above nonlinear equations. 

(a) In the weak dissipation limit ($\gamma \ll g_1, g_2$) and in the odd (o) and even (e) chains, we have
\begin{eqnarray}
    \epsilon_{n,\text{i}}^\text{o} = \frac{(g_1^2 + g_2^2) n^2 \pi^2 \gamma}{ ( g_1  +  g_2 )^2 m^3}, \quad 
    \epsilon_{n,\text{i}}^\text{e} = \frac{2 g_2^2 n^2 \pi^2 \gamma}{ ( g_1 +  g_2 )^2 m^3}.
	\label{eq-weakgamma}
\end{eqnarray}
One can easily check our previous approximation that $\lim_{m\rightarrow \infty} m z_{n,i/r} = 0$. The imaginary part of $\epsilon$ is responsible for the characteristic relaxation time $T^*$ as
\begin{equation}
	T^* = \text{max}(1/(2\epsilon_{n, \text{i}})).
\end{equation}
This result accounts for the multiply time scales during relaxation. When $\gamma \rightarrow 0$, $T^*\rightarrow \infty$, indicating of persistent coherent dynamics. 
Since $\epsilon_{n,\text{i}} \propto \gamma$, it means that in the weak dissipation limit, relaxation is still important 
and can happen in a finite system. Moreover, we find that the odd-even effect is still visible in the long chain limit. 

\begin{figure}
    \centering
    \includegraphics[width=0.48\textwidth]{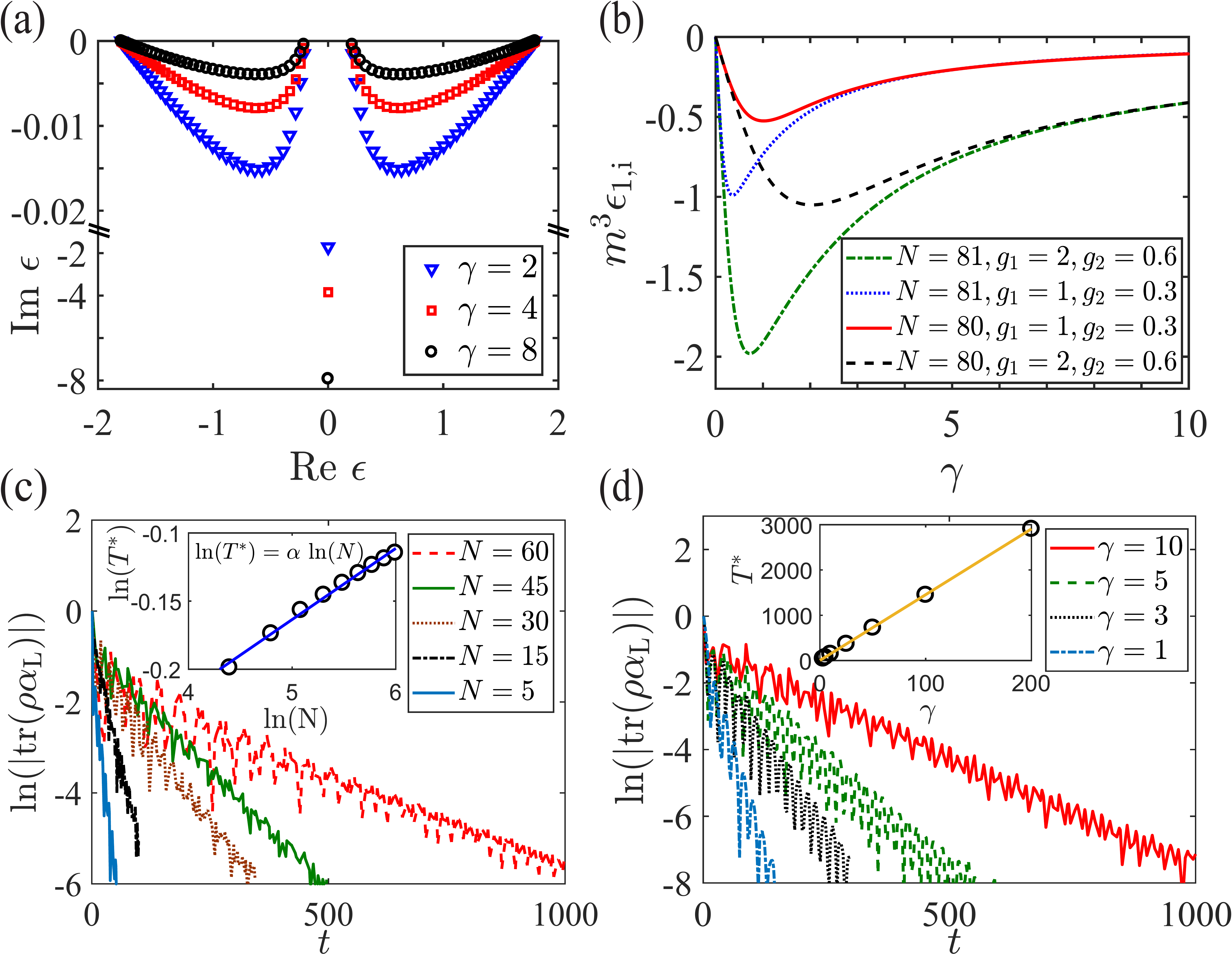}
    \caption{(a) Spectra of the non-Hermitian matrix $H_0+i\Gamma$ for $g_1=1 ,g_2 = 0.8,N=80$ with different dissipation rate $\gamma=2,4,8$.
        (b) $m^3 \epsilon_{1,\text{i}}$ as a function of $\gamma$ and odd-even effect with $N=81$ and $N=80$.
	(c) Long time evolution of $\tr (\rho \alpha_\text{L})$ for different chains with parameters $g_1=1,g_2=0.9,\gamma=10$ .
        Inset shows the scaling law of $\ln T^* = \alpha \ln N$, with fitted parameter $\alpha = 2.3$ in this time window. 
		(d) Long time evolution of $\tr (\rho \alpha_\text{L})$ for different dissipation rates with $g_1=1$, $g_2=0.95$, $N=81$.
        Inset gives the scaling law of $T^* \propto \gamma$. In (c) - (d), $\rho(0)$ is the same as that used in Fig. \ref{fig-fig2} (c) - (d). }
    \label{fig-fig4}
\end{figure}

(b) In the strong dissipation limit, the odd-even effect will vanish, and we find
\begin{equation}
    \epsilon_{n, \text{i}} = \frac{2g_1^2 g_2^2 n^2 \pi^2}{( g_1 +  g_2 )^2 m^3 \gamma}. 
	\label{eq-stronggamma}
\end{equation}
We are surprised to find that in the strong dissipation limit,
the relaxation time $T^* \propto \gamma N^3$, thus it will be prolonged by the dissipation. The crossover between these two cases are determined by 
$\epsilon_{n, \text{i}} = \epsilon_{n,\text{i}}^i$ with $i=$e, o, which yields $\gamma_c = g_1$ in an even chain; and $\gamma_c = \sqrt{2} g_1 g_2 / \sqrt{g_1^2 +g_2^2}$ in an odd chain. 
Thus the strong dissipation regime can be assigned by $\gamma > \gamma_c$. These inflexion points are also numerical verified, which are presented in Fig. \ref{fig-fig4} (b). 
The above solutions are also approximately correct even in a short chain in regarding of the 
fast decay of the imaginary energy according to $\epsilon_{n,	\text{i}} \propto 1/N^3$. 

For the localized edge modes, new decomposition is required. In the weak dissipation limit, the edge modes are not changed by the dissipation and we can make 
perturbation around $\cos(\theta_c)=-{(g_1^2+g_2^2)\over 2g_1 g_2}$, which yields
\begin{equation}
	\epsilon_\text{L} = \epsilon_\text{R} =  -i \gamma (1-\lambda^2),
\end{equation}
for the two modes at the left and right ends. In the strong dissipation limit, the edge modes will be fully localized at the two ends, thus we can assume $\epsilon \simeq -i\gamma$,
and by perturbation about  $\cos(\tilde{\theta}_c)=-{(\gamma^2+g_1^2+g_2^2)\over 2g_1 g_2}$, we have
\begin{equation}
	\epsilon_\text{L} = -i \gamma +i g_2^2/\gamma, \quad \epsilon_\text{R} = -i \gamma +i g_1^2/\gamma,
\end{equation}
in odd chian and
\begin{equation}
      \epsilon_\text{L} = \epsilon_\text{R} = -i \gamma +i g_2^2/\gamma,
\end{equation}
in even chain.
These results are independent of chain length, indicating that in the long chain limit the relaxation time is fully determined by the bulk bands, while the 
edge modes are most vulnerable to dissipation.

\begin{figure}
	\centering
	\includegraphics[width=0.48\textwidth]{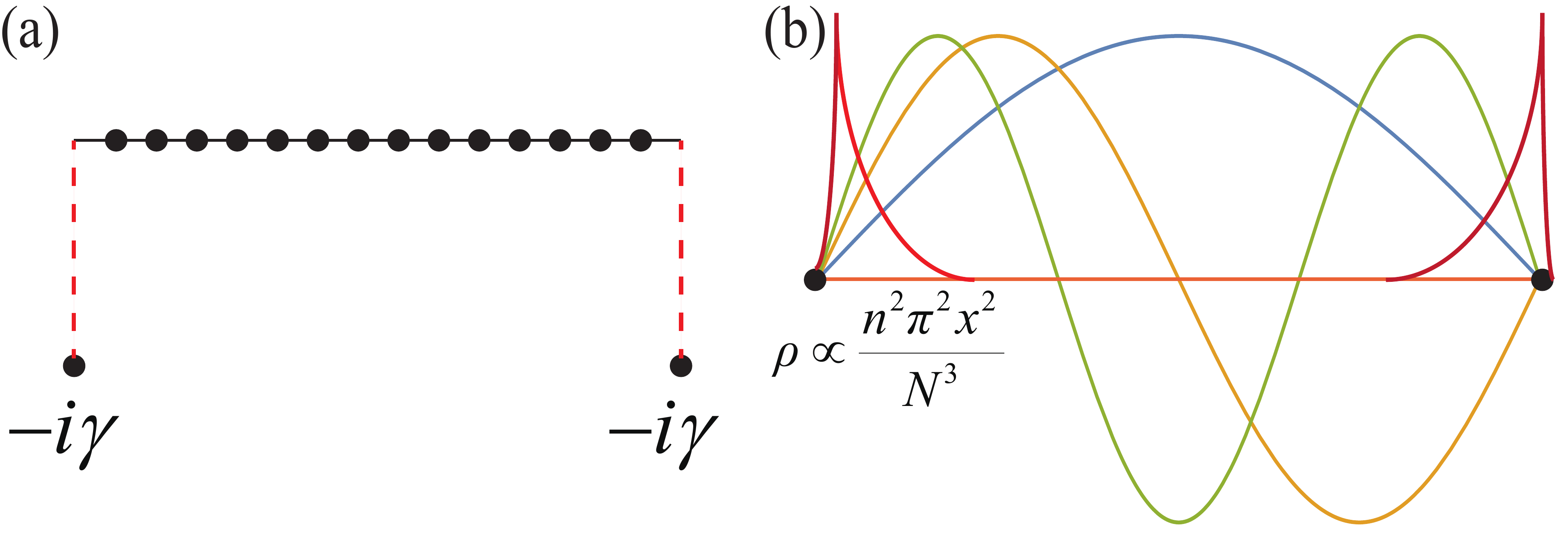}
	\caption{Understanding of the relaxation process induced by boundary dissipation. (a) Effective Hamiltonian with boundary complex potential. 
	(b) Single particle wave functions and their overlaps with the edge dissipation. Here $N = 2m$ or $2m+1$ is the total chain length (see text).}
	\label{fig-fig5}
\end{figure}

{\it Understand of these results}. Let us try to understand these anomalous results using perturbation theory (see Fig. \ref{fig-fig5}). Firstly, in the weak dissipation limit we 
can treat $\Gamma$ as perturbation. Based on the first-order perturbation theory, we find that for the localized edge modes, we have 
$\epsilon_\text{L} = \Im \langle \psi_\text{e} | i\Gamma|\psi_\text{e}\rangle = -i\gamma |\psi_\text{e}(1)|^2 =-i \gamma (1-\lambda^2)$, with wave function 
at the left end as $\psi_\text{e} \sim (1,0,\lambda,0,\lambda^2,\cdots)$ (see Fig. \ref{fig-fig5} (b)). In this case, the dissipation can be independent of total
chain length. For the extended bands in an even chain, the wave function at each site $x$ 
is $\psi_{N \in \text{e}}^n (x \in \text{e}) \propto \sin({n\pi x \over N + {2\lambda \over 1 + \lambda}})$ 
and $\psi_{N \in \text{e}}^n (x \in \text{o}) \propto \sin({n\pi (N+1-x) \over N + {2\lambda \over 1 + \lambda}})$, where $n$ is the band index. In an odd chain, 
$\psi_{N \in \text{o}}^n (x \in \text{e}) \propto \sqrt{\lambda^2+1+2\lambda \cos \theta_n} \sin({(N-x + 1) \over 2} \theta_n)$ and  $\psi_{N \in \text{o}}^n (x \in \text{o}) 
\propto \lambda \sin(({N-x \over 2} + 1)\theta_n) + \sin({L-x \over 2} \theta_n)$, where $\theta_n={2n\pi \over (N+1)}$. 
Then we find $\epsilon_{n,\text{i}} = \langle \psi_N^n|
\Gamma|\psi_N^n\rangle$, which will recover the expression in Eq. \ref{eq-weakgamma}. 
In the strong dissipation limit, we need a different decomposition $H_0 + i\Gamma = \mathcal{H}_0 + i\Gamma + V$,
where $\mathcal{H}_0 = \mathcal{P}_1 H_0 \mathcal{P}_1$ and $V = \mathcal{P}_2 H_0 \mathcal{P}_2 $ ($\mathcal{P}_1$ is the projector into the zero subspace of $i\Gamma$,
while $\mathcal{P}_2$ is orthogonal to $\mathcal{P}_1$). After a few algebra, we find
\begin{equation}
    V=\left(\begin{array}{ccccc}
        0 & ig_{2} & 0 & \cdots & 0\\
        -ig_{2} & 0 &  0 & \cdots & 0\\
        0 & 0 & 0 & \ddots & 0\\
        \vdots & \vdots & \ddots & \ddots & ig_{2}\\
        0 & 0 & 0 & -ig_{2} & 0.
    \end{array}\right),
\end{equation}
which contains only the coupling at sites $2$ and $L-1$, while all the couplings equal to zero exactly. $\mathcal{H}_0$ can be obtained by extracting $V$ from $H_0$ (see Eq. \ref{eq-H0}).
We can treat $V$ as a perturbation, and via the second-order perturbation theory, we have
\begin{equation}
	\epsilon_{n,\text{i}} \propto -\sum_{j=a,b} {\langle \tilde{\psi}_{N}^n |V| j \rangle \langle j| V | \tilde{\psi}_{N}^n\rangle \over (E_n - E_j)},
\end{equation}
where the extended bands $\tilde{\psi}_{N}^n(1)=\tilde{\psi}_{N}^n(N)=0$, $\tilde{\psi}_{N}^n(x) \propto \psi_{N-2}^n(x-1)$, with $x = 2, \dots, N-1$ and the localized modes $| a \rangle = (1,0,0\cdots), | b \rangle = (\cdots 0,0,1)$. This expression yields Eq. \ref{eq-stronggamma} (see the overlap between the bulk modes and the edge dissipation in Fig. \ref{fig-fig5} (b)).

We present the dynamics of $\text{tr}(\rho \alpha_\text{L})$ in Fig. \ref{fig-fig4} (c), where $\alpha_\text{L}$ is the localized edge modes without dissipation. Since strong
dissipation can influence the profile of the edge modes, $\alpha_\text{L}$ is no longer the eigenvector of $H_0 + i\Gamma$. By a linear fitting we find 
\begin{equation}
\ln T^* =\alpha \ln N + \alpha', 
\end{equation}
where $\alpha \sim 2.3$. The exponent of $\alpha = -3$ is not reached due to the finite time window in simulation. This is because we only consider the dissipation of the extended bands 
$\tilde{\psi}_{N}^n$ from its overlap with the edge modes ($\langle \psi_\text{e}| \tilde{\psi}_{N}^n\rangle \neq 0$), thus the dynamics of $\text{tr}(\rho \alpha_\text{L})$ is 
dominated by the extended bands in the long time limit. However, this will not influence the scaling of $T^*$ with respect to $\gamma$ (see Fig. \ref{fig-fig4} (d)), since all the bulk bands 
have the same scaling law $T^* \propto \gamma$.

{\it Conclusion}. Dissipation in the many-body system is a fundamental problem that up to date has not yet been well understood. We explore the dissipation induced
relaxation in a quantum $XY$ model with boundary dissipation, in which the relaxation is characterized by a characteristic time $T^*$. In the long-time
limit it is fully determined by the dynamics of single particle physics. We explore the roles played by edge modes and bulk bands, and their scaling laws with respect to
chain length and dissipation rate. An intuitive picture based on an equivalent non-Hermitian model is proposed. These results may also suggest that the lowest two states protected
by a finite gap width can not serve as quantum memory under dissipation, in consistent with the conclusions in the previous literature \cite{budich2012failure, pedrocchi2015Majorana, 
rainis2012majorana, mazza2013robustness, goldstein2011decay, carmele2015stretched}.

\begin{acknowledgments} {\it Acknowledgement}. This project is supported by  National Key Research and Development Program (No. 2016YFA0301700), National Natural Science Foundation of China(No. 11574294), and the ''Strategic Priority Research Program (B)'' of the Chinese Academy of Sciences (No. XDB01030200). M.G. is also supported by the NSFC ((No. 11774328), the National Youth Thousand Talents Program and the USTC start-up funding.
\end{acknowledgments}


\begin{thebibliography}{59}
\bibitem [{com()}]{complicatetag}%
  \BibitemOpen
  \href@noop {} {}\bibinfo {note} {In a many-body system, the dimension of the
  Hilbert Dim($H$) space increases exponentially with the size of the system.
  However, the dimension of the density matrix for the mixed space in open
  systems should be $\text{Dim}(H)^2$.}\BibitemShut {Stop}%
\bibitem [{dis()}]{dissipationtag}%
  \BibitemOpen
  \href@noop {} {}\bibinfo {note} {This dissipation term written in cannonical
  fermion representation is $\mathcal{D}(\rho) = -\gamma\sum_{j=1,N} (4
  c_j^\dagger c_j \rho c_j^\dagger c_j - 2 c_j^\dagger c_j \rho - 2 \rho
  c_j^\dagger c_j + 2 \rho )$, where $c_1 = \alpha_1 + i \alpha_2$ and $c_N =
  \alpha_{2N-1} + i \alpha_{2N}$.}\BibitemShut {Stop}%
\end{thebibliography}

%
\end{document}